\documentclass[useAMS,usenatbib]{emulateapj}

\shorttitle{Mass Ejection in GRS 1915+105}
\shortauthors{Neilsen et al.}

\begin{document}

\title{Radiation Pressure and Mass Ejection in $\rho$-like States of GRS 1915+105} 

\author{Joseph Neilsen\altaffilmark{1,2,3}, Ronald
  A. Remillard\altaffilmark{1}, Julia C. Lee\altaffilmark{2,3}}
\altaffiltext{1}{MIT Kavli Institute for Astrophysics and Space
  Research, Cambridge, MA 02139; jneilsen@space.mit.edu}
\altaffiltext{2}{Astronomy Department, Harvard University, Cambridge,
  MA 02138}  
\altaffiltext{3}{Harvard-Smithsonian Center for Astrophysics,
  Cambridge, MA 02138}

\begin{abstract}
We present a unifying scenario to address the physical origin of the diversity of X-ray lightcurves within the $\rho$ variability class of the microquasar GRS 1915+105. This `heartbeat' state is characterized by a bright flare that recurs every $\sim50-100$ seconds, but the profile and duration of the flares varies significantly from observation to observation. Based on a comprehensive, phase-resolved study of heartbeats in the \textit{Rossi X-ray Timing Explorer} archive, we demonstrate that very different X-ray lightcurves do not require origins in different accretion processes. Indeed, our detailed comparison of the phase-resolved spectra of a double-peaked oscillation and a single-peaked oscillation shows that different cycles can have basically similar X-ray spectral evolution. We argue that all heartbeat oscillations can be understood as the result of a combination of a thermal-viscous radiation pressure instability, a local Eddington limit in the disk, and a sudden, radiation-pressure-driven evaporation or ejection event in the inner accretion disk. This ejection appears to be a universal, fundamental part of the $\rho$ state, and is largely responsible for a hard X-ray pulse seen in the lightcurve of all cycles. We suggest that the detailed shape of oscillations in the mass accretion rate through the disk is responsible for the phenomenological differences between different $\rho$-type lightcurves, and we discuss how future time-dependent simulations of disk instabilities may provide new insights into the role of radiation pressure in the accretion flow. 
\end{abstract}
                 
\keywords{accretion, accretion disks --- black hole physics --- X-rays:
  individual (GRS 1915+105) ---  X-rays: binaries} 

\section{INTRODUCTION}
\label{sec:intro}
As one of the brightest and most variable sources in the X-ray sky,
the microquasar GRS 1915+105 is an excellent case study for the
investigation of evolving accretion flows. Discovered in 1992 by 
\citeauthor{CastroTirado92}, the black hole has been in outburst for
over 17 years and has become famous for its superluminal jets
\citep{Mirabel94}, its bizarre X-ray variability (\citealt{B00},
hereafter B00; \citealt{K02,Hannikainen05}), and its disk-jet
interactions (\citealt*{FB04,FBG04,NL09,N11a}).

\defcitealias{B00}{B00}
\defcitealias{Massaro10}{M10}

\begin{figure}
\centerline{\includegraphics[angle=270,width=3.3 in]{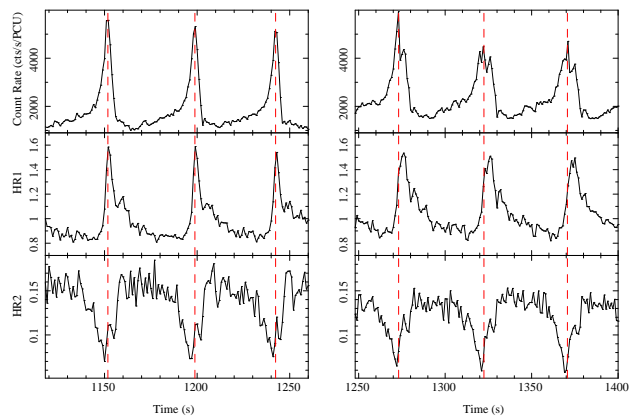}}
\caption{Sample cycles from PCA observations of single-peaked (left)
  and double-peaked (right) oscillations: (top) count rate
  lightcurves, (middle) HR1, and (bottom) HR2. These sample cycles are
  taken from observations 40703-01-07-00 and 60405-01-02-00,
  respectively. In this paper we analyze all \textit{RXTE}
  observations of $\rho$-like cycles; the data here are representative
  of the broad behavior. For reference, we have marked peak count rate
  times (identified by cross-correlation; see Section \ref{sec:obs})
  with dashed vertical lines.} 
\label{fig:lchr}
\end{figure}

Of its many classes of X-ray variability, the best studied are
the $\chi$ state, which produces steady optically thick jets
(see, e.g.\ \citealt{D00,K02}), the $\beta$ state, a wild 30-minute
cycle with discrete ejection events \citep{M98}, and the $\rho$
state, which is affectionately known as the `heartbeat' state for the 
similarity of its lightcurve to an electrocardiogram (see Figure
\ref{fig:lchr}). The $\rho$-type cycle consists of a slow rise
followed by a short bright pulse, repeating with a period of roughly
50 s (\citealt*{TCS97}; \citetalias{B00}).

\begin{figure*}
\centerline{\includegraphics[angle=270,width=\textwidth]{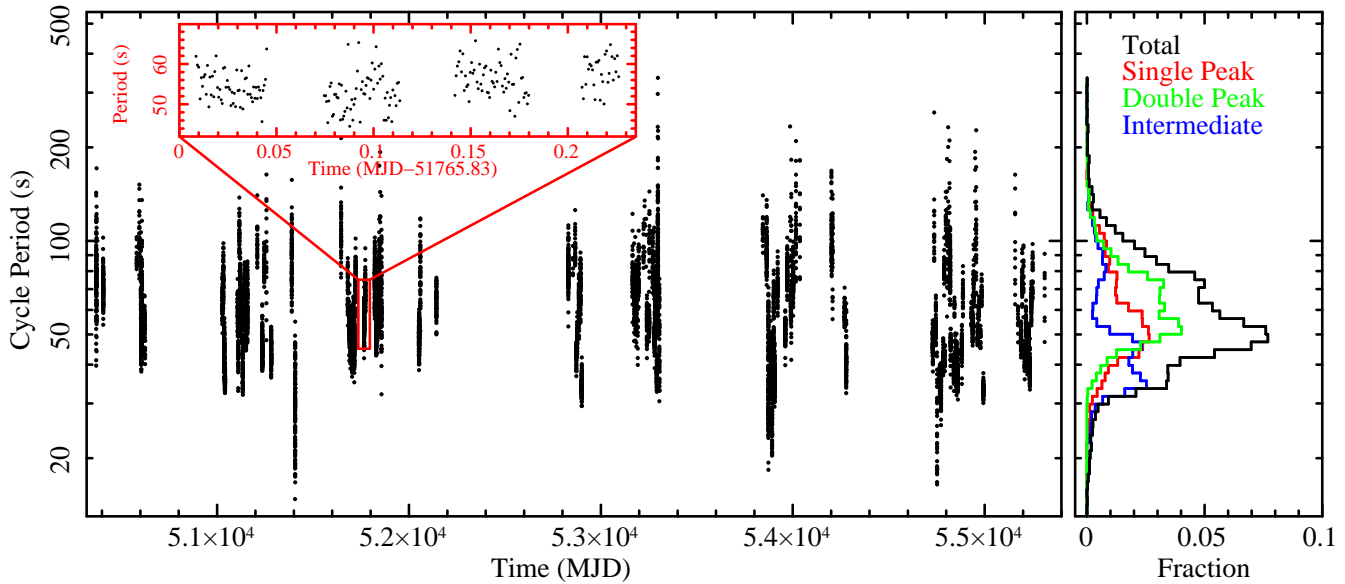}}
\caption{(Left) The measured period of the $\rho$ cycle for all the
  analyzed observations. Periods cluster around 50-60 seconds,
  although there are some excursions as high as 340 seconds. The inset
  shows the drift and scatter that characterize the cycle period in a
  typical observation. (Right) The distribution of periods colored
  according to peak multiplicity.}
\label{fig:period}
\end{figure*}
After nearly two decades of X-ray monitoring of GRS 1915+105, it is
clear that its remarkable diversity of X-ray states is also reflected
in the states themselves. Indeed, even in their original paper on
these variability classes, \citetalias{B00} divide the $\chi$ class
into four sub-categories that differ in hardness and noise
properties. Furthermore, some observations may exhibit characteristics
of multiple states (e.g.\ $\beta$-type variations; \citealt*{N11b}),
and various instances of a single state may even differ in their
large-scale variability patterns. For example, \citeauthor{Massaro10}
(\citeyear{Massaro10}, hereafter M10) recently analyzed a long
\textit{Beppo}SAX observation of GRS 1915+105 in the $\rho$ state and
reported significant changes in the lightcurve shape: the number of
pulses or peaks per cycle ranged from one to at least four. These
pulses can be separated by 10 seconds or more, and the later peaks are
typically harder \citep{TCS97,Paul98}.

In the hopes of uncovering direct evidence for the origin of these
luminous $\rho$-type pulses, we performed the first joint
\textit{Chandra/RXTE} phase-resolved spectral analysis of an
observation of GRS 1915+105 in the $\rho$ state \citep{N11a}. Using
the \textit{Chandra} gratings \citep{C05}, we showed for the first
time that changes in the broadband X-ray spectrum drive physical
changes in the accretion disk wind on time scales as short as 5
seconds, and that this wind is sufficiently massive to cause state
transitions in the disk. Based on \textit{RXTE} spectra, we argued 
that radiation pressure plays several key roles in the accretion disk,
from driving the observed limit cycle via the Lightman-Eardley
instability (a.k.a.\ radiation pressure instability;
\citealt{Lightman74,B97b,J00}) to literally pushing the inner edge of 
the accretion disk away from the black hole (a local Eddington limit;
\citealt{Fukue04,Heinzeller07}; \citealt*{Lin09}) until the global
Eddington limit is reached and the disk is evaporated or ejected at
the maximum accretion rate (see also \citealt{JC05}). For this last
point, we argued that the second peak in our double-peaked lightcurves
could be explained as bremsstrahlung emitted when the ejected gas
collides with the corona. 

But what is the significance of these results when some instances of
the $\rho$ state only have a single peak \citepalias{Massaro10}? That
is, if there is no ``second peak'' in the lightcurve, is mass ejection
still required to explain the X-ray oscillation? Are we to understand
different variations of the $\rho$-like cycles as driven by the same
physical processes or as producing similar lightcurves by coincidence?
In this paper, we begin to address the remarkable diversity
\textit{within} the variability classes of GRS 1915+105, with a
comprehensive phase-resolved study of all the heartbeat-like states in
the \textit{RXTE} archive.

We describe the observations, data analysis, and phase-folding in
Section 2. In Section 3, we perform a detailed comparison of the
timing and spectral behavior of single- and double-peaked $\rho$
states. In Section 4, we interpret our results as clear evidence that
a single physical scenario may produce a wide variety of lightcurves,
and we consider the exact origin of the variations observed by
\citet{Massaro10}. 

\begin{figure*}
\centerline{\includegraphics[width=\textwidth]{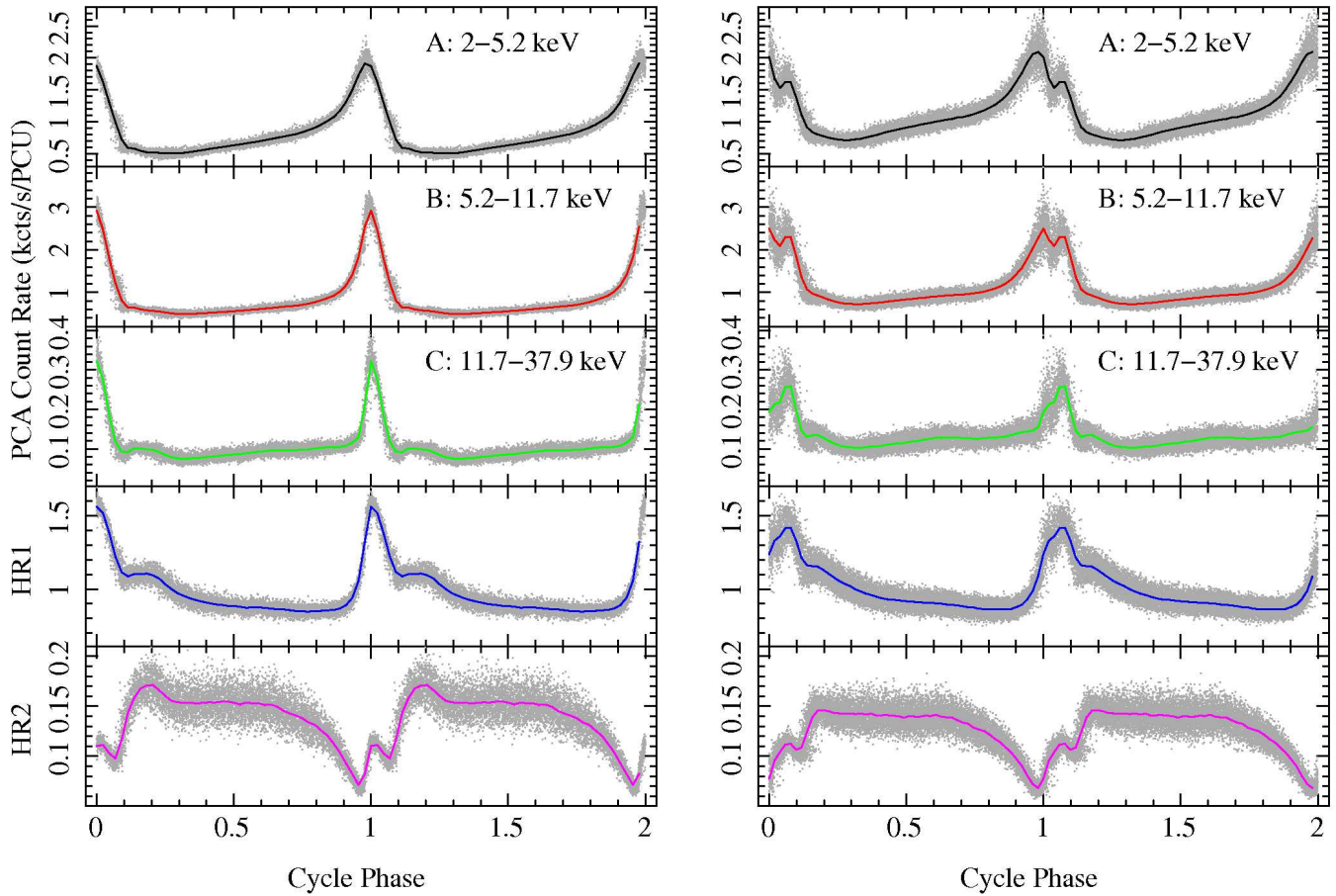}}
\caption{Average phase-folded lightcurves and hardness ratios for the
  single- and double-peaked oscillations depicted in Figure
  \ref{fig:lchr}, i.e.\ \textit{RXTE} observations observations
  40703-01-07-00 and 60405-01-02-00, respectively. Two cycles are
  shown for clarity. The smooth curves represent the average
  lightcurve profiles, and the data are plotted in gray (204 cycles on
  the left, 265 cycles on the right). It is obvious
  that despite the scatter and variations in the period, the phase
  profile of the $\rho$ oscillation is remarkably constant from cycle
  to cycle. Note that the energy ranges plotted here differ from
  \citet{N11a} because we are using new, normalized lightcurves
  (Section \ref{sec:obs}).}
\label{fig:philc}
\end{figure*}

\section{OBSERVATIONS AND PHASE FOLDING}
\label{sec:obs}
Since its launch in 1995 \citep{J96}, RXTE has observed GRS 1915+105
roughly twice a week, for a total of over 1600 pointings as of the
original writing of this paper. In this paper, we will focus
exclusively on the data from the Proportional Counter Array (PCA),
which covers approximately 2--60 keV. We extract 1-second normalized
lightcurves from the binned data modes in three energy channels:
$A\equiv2-5$ keV, $B\equiv5.2-12$ keV, and $C\equiv12-45$ keV. The
lightcurves are normalized using the PCA lightcurve of the Crab
nebula. That is, for each PCU, we renormalize the mission-long Crab
lightcurve to count rate values 1100, 1140, and 330 counts per second
for channels $A,~B,$ and $C,$ respectively. This roughly corresponds
to the raw count rate of PCU2 around MJD 52000. We then apply these
same normalizations to the GRS 1915+105 count rates so that the
intensity scale is the same for all observations. We also produce two
hardness ratios: HR1$=B/A$ and HR2=$C/B.$

In order to identify $\rho$ state observations, we examined 1-second
lightcurves and color-color diagrams (CD) for all observations of GRS 
1915+105. We selected observations with regular or quasi-regular
bursts of the appropriate shape, duration, and CD by visual
inspection. This task can be rather tricky since there are variations
within the heartbeat state (Section \ref{sec:intro}) as well as strong
resemblances between the $\rho$ class and some portions of the $\nu$
and $\kappa$ classes; our analysis may include some cycles from these
classes. However, the fact that our phase-folding method can be used 
for all these quasi-regular cycles is indicative of the physical
robustness of our results. We show samples of the lightcurves and
hardness ratios for two typical (long) observations in Figure
\ref{fig:lchr} (\textit{RXTE} observations 40703-01-02-00 and
60405-01-02-00 on the left and right, respectively). The oscillation
is obvious in all panels. Inspection of these data and our ensemble of
lightcurves reveals the same variations noted by
\citetalias{Massaro10}, namely that the number of strong peaks in the
lightcurve changes over time scales of days or longer, as do the
durations of the pulses and the time delays between them. As we will
see later that there is likely a continuum of lightcurve shapes, we
will not devote any time here to detailed phenomenological
categorizations of these profiles.

In \citet{N11a}, we showed that we can very accurately
characterize the physics of this oscillation by phase-folding the
individual cycles, which takes into account the variable oscillation
period. In other words, we stretch or compress and then combine all
the individual cycles in a given observation, and study the spectral
conditions at each part of the cycle. We determine the start times of
each cycle by means of an iterative cross correlation method that is
described in more detail in \citet{N11a}. Briefly, we use
cross-correlations to identify the main peak in the count rate for
each cycle, and then define the times of peak count rate as phase
$\phi\equiv0.$ With this analysis, we measure 10068 
peak times in 242 observations. For reference, we plot the resulting
cycle periods (i.e.\ the distance between successive times of
$\phi=0$) in Figure \ref{fig:period}. It is evident from this figure
that typical periods are $\lesssim100$ seconds, although some single
cycles may last over 300 seconds. Within a given observation, there is
a fair amount ($\gtrsim10\%$) of cycle-to-cycle scatter in the period,
but the period is usually very stable around the mean (see inset,
Fig.\ \ref{fig:period}).

Once we have defined a phase ephemeris for all 242 observations, we
create phase-folded lightcurves for each, and then we identify them as
single-peaked, double-peaked, or as intermediate cases. Observations
may be classified as intermediate if there is substantial noise
(e.g.\ if the observation only includes a few cycles) or if the cycles
have unusual shapes or strong period variability (like $\kappa$-type
cycles), so that it is difficult to reliably distinguish two
individual pulses from a single pulse. Of the 242 observations
included here, we classify 101 as having double-peaked cycles, 77 as
single-peaked, and the remaining 64 as intermediate cases. Because
there is an inherent difficulty in deciding whether structure in the
lightcurve is noise or a physical peak, there is an uncertainty in
these numbers of about 10--15 (estimated by repeating the
classification). This uncertainty has a negligible effect on the
numbers presented in this section and no effect on the subsequent
spectroscopy our physical interpretation.

There are some statistical differences in the period between cycles
with different numbers of peaks. The right panel of Figure
\ref{fig:period} shows that single- and double-peaked cycles have
similar periods, at $61\pm23$ s and $64\pm18$ s, respectively, while
intermediate cycles are generally shorter, with an overall mean of
$52\pm27$ s. The uncertainties represent the sample standard
deviations; the standard errors of the mean are $\lesssim0.5$
s. At very short periods, the intermediate cycles show a bimodal
period distribution, which is reminiscent of the $\kappa$ state
\citepalias{B00} or the irregular bursts studied by
\citet{Yadav99}. However, we will see that these intermediate cases
exhibit similar spectral behavior to single- and double-peaked cycle
types, so they are unlikely to represent a completely different
physical phenomenon.

Figure \ref{fig:philc} shows the phase-folded lightcurves for the
single- and double-peaked observations illustrated in Figure
\ref{fig:lchr}. The average phase-folded lightcurve profiles, shown as
solid lines, are superimposed on the individual cycles from these
same observations shown as gray points. There are a total of
204 cycles and 265 cycles on the left and right, respectively. The
scatter is limited and it is clear that within an observation the
individual cycles have the same profile. On the other hand, it is also
clear that heartbeats can differ substantially from one observation to
the next (i.e.\ in terms of the number of peaks or pulses). 
Building on our prior study of double-peaked cycles \citep{N11a},
we focus in this paper on a comparison between single-peaked cycles
and double-peaked cycles, and the question: can these two cycle types
be produced by the same mechanism? In other words, does the single
pulse correspond to the first (soft) pulse of a double-peaked cycle,
the second (hard) pulse,
a combination of the two pulses, or is it completely different?

\section{THE SINGLE PULSE}
\label{sec:peak}
In this section, based on timing analysis and broadband spectroscopy,
we develop a case that the single pulse in single-peaked cycles
corresponds to the second, hard pulse in double-peaked cycles. As
such, our results will indicate that the physical processes
responsible for the production of this hard pulse are active in all
$\rho$-like cycles. In other words, these processes are fundamental to 
the heartbeat state.

\begin{figure}
\centerline{\includegraphics[width=3.3 in]{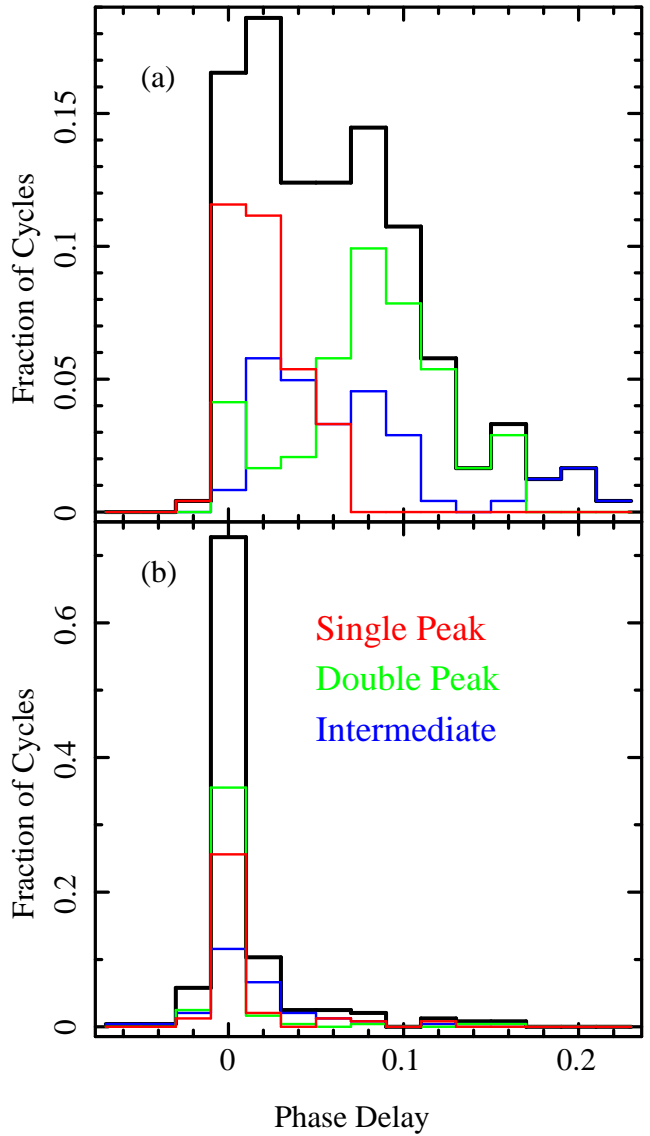}}
\caption{Distribution of phase delays between maxima in (a) the $A$
  and $C$ bands and (b) the $C$ band and HR1, for all observations
  analyzed in this paper. The top panel shows that single-peaked
  cycles generally have a small delay between their soft and hard
  maxima, but double-peaked oscillations have a much longer delay. In
  this respect, the intermediate cases sometimes mimic single-pulse
  profiles and sometimes behave like double-peaked cycles. The bottom
  panel demonstrates that there is no significant difference between
  any cycle types in the delay between the maximum in HR1 and the
  maximum in the hard X-ray lightcurve, although HR1 typically peaks a
  little later than the $C$-band lightcurve itself.}
\label{fig:delay}
\end{figure}
\subsection{Folded Lightcurves and Timing Analysis}
\label{sec:timing}
\begin{figure}
\centerline{\includegraphics[width=3.3 in]{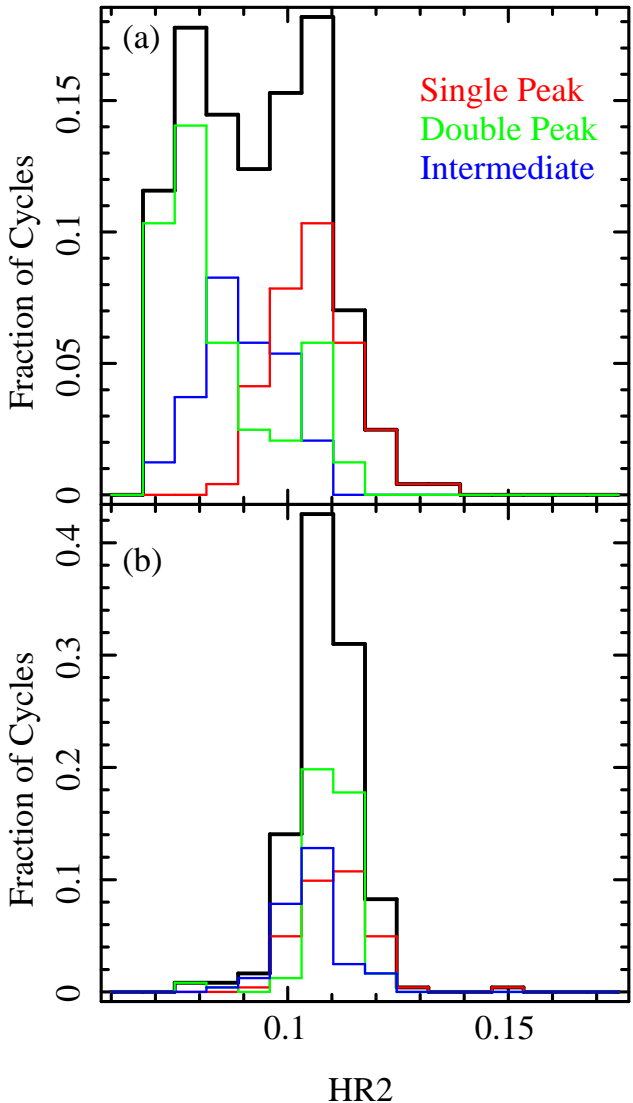}}
\caption{The distribution of HR2 at two phases of the heartbeat cycle
  for all observations analyzed in this paper: (top) $\phi=0$, and
  (bottom) at the maximum of the $C$-band lightcurve. The top panel
  shows that double-peaked cycles are typically softer than
  single-peaked cycles at their respective times of $\phi=0,$ while
  the bottom panel shows that both types of cycle have very similar
  hardness near the maximum of the $C$-band lightcurve. Thus the
  single pulse may correspond to the \textit{hard (second)} pulse of
  double-peaked cycles.} 
\label{fig:hr}
\end{figure}
The first indications that the single pulse is analogous to the hard
pulse come from the lightcurves and phase-folded lightcurves in
Figures \ref{fig:lchr} and \ref{fig:philc}. In these plots, HR2 rises
sharply from its minimum value around $\phi=0$, giving the appearance
of a local maximum. The local maximum in HR2 coincides with the
maximum value of HR1 in both single-peaked and double-peaked
cycles. This phase of ``maximum spectral hardness'' occurs during the
main pulse in single-peaked cycles, while it is very close to the
second pulse when there are two peaks in the cycle.

An alternative indication of this same phenomenon is the delay between
various maxima in the hardness ratios and in the hard and soft X-ray
lightcurves. For example, consider that in a double-peaked cycle, HR1
peaks at the same time as the $C$-band lightcurve, but after the $A$
band, while all three quantities are synchronized in single-peaked
observations. 

The delays between these maxima are easily quantified from the
phase-folded lightcurves, and we present the phase delay distributions
in Figure \ref{fig:delay}. It should come as no surprise that
double-peaked oscillations exhibit a significant delay between the
hard and soft X-ray pulses (top panel):
$\Delta\phi_{C-A}=0.08\pm0.05$, 
where the uncertainty is the sample standard deviation. As the average
period of these oscillations is $\sim64$ seconds, we find that the
hard pulse typically follows the soft pulse by $\sim4.9$ 
seconds. For
single-peaked cycles, the delay between the peaks in the $A$ and $C$
bands is much smaller (consistent with zero):
$\Delta\phi_{C-A}=0.02\pm0.02.$ 
The intermediate
cases show a bimodal distribution, apparently including mixed
contributions from single-peaked and double-peaked cycles. As
suggested above, all cycle types exhibit a very short delay between
the maximum of HR1 and the hard X-ray pulse. This delay is
$\Delta\phi_{\rm HR1-\it{C}}=0.02\pm0.04$ for single-peaked cycles,
$0.01\pm0.04$ for double-peaked cycles, and $0.01\pm0.03$ for
intermediate cases.  
\begin{figure*}
\centerline{\includegraphics[width=\textwidth]{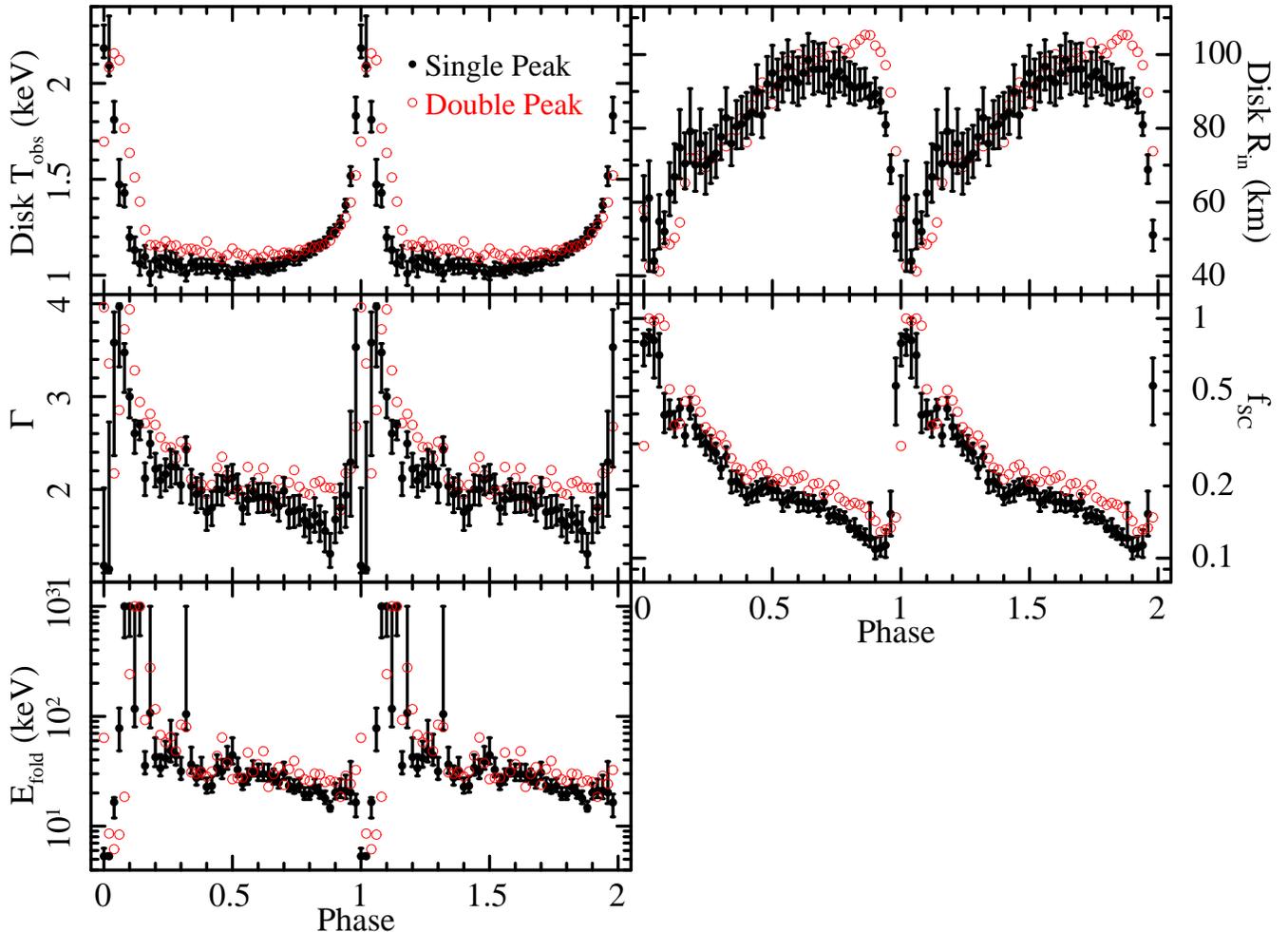}}
\caption{Model fit parameters and their 90\% confidence limits as a
  function of phase for the PCA spectra of the single-peaked heartbeat
  oscillation (filled black points; \textit{RXTE} observation
  40703-01-07-00). For comparison, we overplot our parameters for a
  double-peaked cycle (open red circles; \citealt{N11a}, \textit{RXTE}
  observation 60405-01-02-00). Two cycles are shown for clarity.
  Since   the oscillation periods are near 50 seconds, each phase bin 
  corresponds to roughly one second of real time. The fits are
  extremely similar; in conjunction with the phase shift between the
  two parameter sets, this result implies that the pulse in
  single-peaked cycles physically corresponds to the hard second pulse
  in double-peaked cycles.} 
\label{fig:spec}
\end{figure*}

Also relevant are the values of the hardness ratios themselves. In
Figure \ref{fig:hr} we present the distribution of HR2 at $\phi=0,$
when the count rate is at a maximum (top), and at the maximum of the
$C$-band lightcurve (bottom). In general, we see that at $\phi=0$,
double-peaked oscillations are softer than cycles with a single
peak (intermediate cases lie in between the two). This is to be
expected if the single pulse corresponds to the hard pulse of a
double-peaked cycle. That is, when there are two peaks, our
cross-correlation method picks out the brighter, softer first pulse
for $\phi=0$; in the absence of the soft pulse (i.e.\ one pulse only),
$\phi=0$ coincides with the hard pulse. The bottom panel of Figure
\ref{fig:hr} confirms this conjecture: all cycles have very similar
HR2 when the $C$-band flux is maximized. This suggests that a common  
light source produces a pulse of hard X-rays in every cycle.

Again, the relative timing of the X-ray flux
and spectral hardness ratios indicates that the intermediate cases
represent a continuum of profiles between single- and double-peaked
oscillations. A completely different physical process might be less
likely to produce the same timing signatures, so we conclude that at 
1-second resolution, the variations in the cycle profile can be
explained in the context of two main pulses, one soft and one hard,
whose relative strengths, duration, and separation can vary from
observation to observation. In Section \ref{sec:spec}, we employ
phase-resolved spectra to unify these pulses in a single physical
framework. 

\subsection{Phase-Resolved Spectroscopy}
\label{sec:spec}
In \citet{N11a}, we showed that in a double-peaked cycle
(\textit{RXTE} ObsID 60405-01-02-00), the X-ray spectrum of the hard
pulse was described very well as the sum of two major components: a
disk blackbody and a power-law type component, possibly bremsstrahlung
or Comptonized disk emission. Statistically, we could not distinguish
between different Comptonization models and bremsstrahlung emission,
but all models required a sudden decrease in the electron temperature
associated with this component during the hard pulse. 

In this section, we ask whether an observation of a single-peaked
cycle can be described by the same model with similar phase
dependence; we return to the physical interpretation of this model in
Section \ref{sec:discuss}. We perform phase-resolved spectroscopy of
the single-peaked observation represented in Figures \ref{fig:lchr}
and \ref{fig:philc}, \textit{RXTE} ObsID 40703-01-07-00, and compare
to our results on the double-peaked cycle in \citet{N11a}. We
implicitly assume that these two observations are representative of
single- and double-peaked heartbeats. This assumption is justified by
our results in Section \ref{sec:timing}, which demonstrate clear
similarities between different observations of a given type of cycle.

Our spectral model  consists of interstellar absorption by cold gas
({\tt tbabs}; \citealt*{Wilms00}), a hot disk ({\tt ezdiskbb};
\citealt{Zimmerman05}) convolved through a scattering kernel 
({\tt simpl}; \citealt{Steiner09a}), a Gaussian emission line at 6.4
keV, and a high-energy cutoff ({\tt highecut}, with functional form
$\exp(-E/E_{\rm fold})$ for $E>E_{\rm cut}$; we set $E_{\rm  cut}=0$). 
{\tt simpl} takes a seed spectrum and scatters a fraction $f_{\rm SC}$
of the source photons into a power law, approximating the
high-temperature, low optical depth regime of Comptonization. We use
{\tt highecut} to account for curvature in the hard X-ray spectrum,
and we prefer {\tt simpl} to {\tt powerlaw} because it conserves
photons and includes a physically-realistic rollover at low
energies. Our method is precisely the same as that reported in
\citet{N11a}: we extract phase-resolved spectra and responses and,
applying the model described above, we fit the emission from 3--45
keV. All spectral fitting is done in ISIS \citep{HD00,Houck02}. We
assume a distance and inclination of $D = 11.2$ kpc and $i =
66^{\circ}$ \citep{Fender99}; we fix $N_{\rm H} = 5\times10^{22}$
cm$^{-2}$ (\citealt{L02} and references therein). Overall, the model
provides an excellent description of the data, with a reduced
$\chi^{2}/\nu=1847.6/2149=0.86.$ 

The resulting fit parameter values for our single-peaked oscillation
are shown in black in Figure \ref{fig:spec}. The maximum observed
temperature in the disk, $T_{\rm obs},$ is mostly constant near 1.1
keV throughout the cycle but spikes to $\sim2.2$ keV at $\phi=0.$
While the disk temperature hovers around 1.1 keV, the inner disk
radius $R_{\rm in}$ rises steadily from $\sim60$ km to $\sim95$ km,
but then turns over and plummets quickly to $\sim40$ km (we use a
color-correction factor $f=1.9,$ appropriate for high luminosity;
J.\ Steiner, private communication). The {\tt simpl} photon index
$\Gamma$, the scattering fraction $f_{\rm SC},$ and the high-energy
cutoff $E_{\rm fold}$ all decrease steadily for most of the cycle,
from roughly 4 to 1.5, 0.9 to 0.1, and $\gtrsim$ 500 keV to 30 keV,
respectively. Near $\phi=0,~\Gamma$ and  $f_{\rm SC}$ rise sharply,
but $E_{\rm fold}$ (i.e.\ the electron temperature) dips to 5
keV. During this brief dip, there is also weak evidence of a
coincident dip in $\Gamma.$

For comparison, we overplot our \citep{N11a} measured parameters for
the double-peaked oscillation. Given the differences in the
lightcurves (Figure \ref{fig:lchr}), the similarity of the spectral
fits is striking. 
The most important difference is a clear phase shift: a given feature
in a given parameter occurs later in phase in double-peaked cycles. In
terms of the parameters themselves, the double-peaked cycle has a
slightly higher and flatter baseline in $T_{\rm obs}$, and the spike
is smaller; the disk radius increases for a longer interval relative
to the single-peaked cycle. The double-peaked cycle also has a
slightly larger photon index, scattering fraction, and electron
temperature than the single-peaked oscillation. 

Two points from our spectral fits deserve special consideration: 
\begin{enumerate}
\item It is remarkable that in these two very different cycles, the
  measured radius of the disk plummets sharply and comes to rest
  (albeit briefly) at the same value of $R_{\rm in}\sim40$ km. The
  possibility of a ``touchdown'' radius is especially interesting
  given that 40 km is the radius where one expects the peak emission
  from a thin disk around a maximally-spinning 14 M$_{\odot}$ black
  hole \citep{McClintockShafee06}.
\item Both single-peaked and double-peaked oscillations are marked by
  a short interval where the scattering fraction approaches 1 and
  $E_{\rm fold}$ drops to $\sim5$ keV, indicating the sudden
  appearance of an optically-thick population of relatively cool
  electrons. In double-peaked cycles, this interval corresponds to the 
  hard pulse. In single-peaked cycles, it coincides with the main peak
  in the lightcurve at $\phi=0.$ This connection confirms the 
  identification of the single pulse as the hard pulse.
\end{enumerate}
On the whole, it is abundantly clear that the same spectral processes
are operating in both of these cycles. 
\section{DISCUSSION}
\label{sec:discuss}
In Section \ref{sec:peak}, our phase-resolved timing and spectral
analysis demonstrated that single-peaked heartbeats should be
interpreted as double-peaked cycles with a weak soft pulse, while
the brief pulse of hard X-rays is truly fundamental to the $\rho$
cycle. In this section, we consider the origin of these pulses and the
implications of their relative variations.

In our detailed phase-resolved spectral analysis of a double-peaked
oscillation \citep{N11a}, we explored two Comptonization
models ({\tt simpl} and {\tt nthcomp}; \citealt{Zdziarski96,Zycki99})
and also bremsstrahlung as descriptors of the broadband X-ray 
emission in the heartbeat state. While these models differ in their
physical interpretations, they implied very similar behavior
in the inner accretion disk, and all three models required the sudden
appearance of a new population of electrons with a temperature of
$\sim5$ keV during the hard pulse. We argued that this was probably
due to the sudden ejection of matter from the disk at very high
luminosity (see \citealt{JC05} for a theoretical basis). 

To produce the waves in $\dot{M}$ that cause the system to exhibit
these changes on a cyclic basis, we invoked a global Eddington
instability (\citealt{Lightman74,N11a}). This instability, a class of
thermal-viscous instability (\citealt{TCS97}; \citealt*{Nayakshin00}),
has also been called the radiation pressure instability
\citep*[RPI;][]{J00,JC05}, a term we adopt in what follows. We also found
a long interval of increasing disk radius at constant temperature,
which we argued was an indication of a local Eddington limit in the
disk \citep{Fukue04,Heinzeller07,Lin09}. Local Eddington effects can
occur in geometrically thin disks when the local vertical component of
radiation pressure becomes sufficiently strong to overcome gravity,
disrupting the disk interior to some critical radius. In short, when
the RPI raises the local accretion rate, the disk responds via the
local Eddington process: it maintains a constant temperature but
increases its inner radius as it expels a ring of gas at $R_{\rm in}$ 
\citep{Fukue04,Heinzeller07,Lin09}. These phenomena (mass ejection and
the local Eddington process) reflect the critical importance of
radiation pressure in the accretion dynamics of double-peaked
heartbeat states.  

Furthermore, in the single-peaked cycle we have studied here, the same
two effects (i.e.\ mass ejection and disk evolution via a local
Eddington limit) also dominate the spectral evolution. Thus we believe
that a single unified model can successfully describe the various
lightcurves of $\rho$-like states, and we identify radiation pressure
as the driver of this unified model. From the oscillating mass 
accretion rate (the radiation pressure instability) to the slow
growth of the disk radius (the response of a luminous thin disk to the
rising accretion rate) to the ejection or evaporation of the inner
disk (a consequence of the sudden final influx of matter), radiation
pressure mediates all major aspects of the $\rho$ cycle.

\begin{figure}
\centerline{\includegraphics[width=3.3 in]{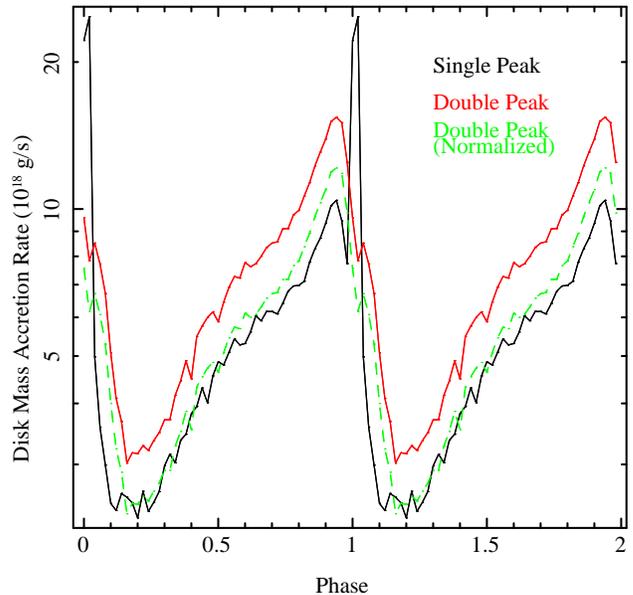}}
\caption{Accretion rates versus phase implied by our fits to the
  phase-resolved PCA spectra of single- (black) and double-peaked
  heartbeats (red). For direct comparison, we also overplot $\dot{M}$
  for the double-peaked cycle, normalized to the mean value for the
  single-peaked cycle (green dashed line). In double-peaked cycles,
  the accretion rate is higher and the wave of matter is sustained for
  longer; in single-peaked cycles, the overall accretion rate is lower
  and spikes sharply around $\phi=0.$ We discuss in Section
  \ref{sec:discuss} how the differences in the $\dot{M}$ profile may
  be responsible for the different X-ray lightcurves.}
\label{fig:mdot}
\end{figure}

If a common set of accretion processes can describe the range of
$\rho$-like cycles seen in the \textit{RXTE} archive, we must also
address the question: why are some oscillations different than others?
If the hard pulse is always observed in the heartbeat state, where
does the soft pulse go? To answer this important question, we
look to the comparison of our spectral parameters in Figure
\ref{fig:spec}. The salient features of the phase evolution of the
disk are the plunge of $R_{\rm in}$ towards the black hole and the
rapid heating of the inner disk. Ultimately, it is the timing of this
plunge relative to the spike in the disk temperature that determines
the cycle profile and the relative strengths of the two pulses. 

But examining the evolution of $R_{\rm in}$ and $T_{\rm obs}$ in
light of the role of radiation pressure in the heartbeat state, we see
two puzzles that merit further consideration. First, why are the
changes in the disk so catastrophic? In principle, the local Eddington
process is reversible, so that $R_{\rm in}$ could also
\textit{decrease} slowly at roughly constant temperature
(e.g.\ reversible changes tied to the local Eddington limit appear to
operate along the flaring branch of Z-sources; \citealt{Lin09}). But
in the heartbeat state of GRS 1915+105, the symmetry appears to be
broken: the disk expands and contracts along vastly different paths in
the $R_{\rm in}-T_{\rm obs}$ plane. It is unclear whether this
asymmetry is related to mass ejection, a steep gradient in the
accretion rate, reaching some energetic limit in the disk, or some
other unknown factors, but it is clear that there is significantly
more to the behavior of the disk than the local Eddington process.

This brings us to the second puzzle: what causes the local Eddington
effect to stop acting on the disk? In \citet{N11a}, we found that in a
double-peaked cycle, $R_{\rm in}$ grows at roughly constant
temperature until $\phi\sim0.9,$ when the catastrophic changes take 
over to begin a new cycle in the disk. But for the single-peaked cycle
shown in Figure \ref{fig:spec}, $R_{\rm in}$ stops growing and actually 
decreases slightly from $\phi\sim0.75-0.9$. This momentary plateau in
$R_{\rm in}$ signals the end of the local Eddington process, but here
it occurs almost ten seconds before the catastrophic changes in the
disk. In other words, the plunge of $R_{\rm in}$ does not cause the
end of the local Eddington effect (although they may coincide, as in 
double-peaked cycles; \citealt{N11a}). Instead, the end may be
controlled by changes in the accretion rate. From our fits, this
puzzling phase interval shows a steady $R_{\rm in}$ plus an increasing
$T_{\rm obs},$ indicating an increasing accretion rate. So it may be
that a particular range in $\dot{M}$ is required to maintain the local
Eddington effect. But without a clear theoretical answer to these
questions, we consider a number of intriguing possibilities for the
disk evolution.

First, we can take the results of our phase-resolved spectroscopy at
face value and use our disk parameters to infer the disk mass
accretion rates for single- and double-peaked oscillations (Figure
\ref{fig:mdot}). We calculate the accretion rate using Equation 4 of
\citet{Zimmerman05}. To facilitate a visual comparison, we also show the 
double-peaked $\dot{M}$ vertically normalized to match the mean value
for the single-peaked cycle (green dashed line). The accretion rate is
higher on average in the double-peaked cycle ($\dot{M}\approx7.4\times 
10^{18}$ g s$^{-1}$ versus $\dot{M}\approx5.9\times10^{18}$ g
s$^{-1}$), but the $\dot{M}$ wave is wider and has a smaller absolute
amplitude. As an aside, this higher accretion rate implies a larger
unstable region in the disk, which could explain why the cycle period
is slightly larger for double-peaked cycles
(e.g.\ \citealt{Nayakshin00}). Considering the last seconds
before/during the collapse of the disk, we see that $\dot{M}$ is
sharply peaked in the single-peaked cycle, so that for a fixed
phase-averaged accretion rate (i.e.\ comparing the normalized
accretion rates), more matter accretes earlier in double-peaked
oscillations.

But this explanation is only valid to the extent that the accretion
disk dominates both the mass flux and the X-ray emission. If there are
outflows from very small radii\footnote{\textit{Chandra} observations
of GRS 1915+105 indicate the presence of winds, but these are
typically driven from the outer disk (\citealt{M08,NL09,N11a,N11b}).} 
or if there is a second component, e.g.\ the corona, with a non-zero
accretion rate or some difference radiative efficiency, then Figure
\ref{fig:mdot} is only an approximation. We note that our accretion
rates are calculated from the disk \textit{prior to} Compton
scattering. For this reason, the contribution of the corona could be
further involved if there is significant mass transfer between it and
the disk, as suggested by \citet{JC05} and verified by our fits here
and in \citet{N11a}. \citet{JC05} consider several different
prescriptions for mass transfer between the accretion disk and the
corona (i.e.\ dependent on the luminosity or the accretion rate in the
disk). It could be that single- and double-peaked oscillations
transfer mass differently, and that this controls the evolution of the
local Eddington process in the disk. As the corona has a stabilizing
effect on the disk \citep{JC05}, it may also be that the coronal
geometry differs between different types of cycles, leading to
different plunge behavior. 

Figure \ref{fig:mdot} is also only accurate insofar as the disk can be
approximated as having a constant radiative efficiency, optical depth,
spectral hardening, and a constant/uniform accretion rate (in a given
phase bin) in the inner disk. The accuracy of these approximations
must be determined from simulations that calculate and fit spectra at
each phase of the cycle and compare to the physical behavior of the
accretion disk in the simulations. Since \citet{Nayakshin00} and
\citet{JC05} invoke strong variations in the disk viscosity and
surface density, it may be that our approximations here are
insufficient for a complete understanding of the evolution of the
disk. On the other hand, it seems clear that the phase dependence of
the accretion rate (whether as shown in Figure \ref{fig:mdot} or
otherwise) is probably the primary factor in determining the strength
and duration of the X-ray pulses, and may in some yet-to-be-determined
way influence the duration of the local Eddington
effect. \citet{Nayakshin00} note that the cycle period and profile
depend on the mass accretion rate in a non-linear way, and it would be
very interesting to see if future simulations including a local
Eddington effect and mass transfer between the disk and corona
\citep[after][]{JC05} could reproduce the peak structure of the
observed cycles.

The need for future theoretical studies is also highlighted by a
recent spectral study of a long \textit{Beppo}SAX observation of the
heartbeat state \citep{Mineo12}. Dividing the cycles into 5
characteristic intervals, \citet{Mineo12} followed the disk and corona
during the cycle, and they observed behavior consistent with our
present work and \citet{N11a}: the disk radius increases at constant  
temperature and the coronal electron temperature drops when the disk
plunges inwards and the disk temperature spikes (i.e.\ in the hard
pulse). However, their results suggested a \textit{decrease} in the
coronal optical depth during the hard pulse, which they interpreted as
a condensation of the corona (in contrast to the sharp increase in the
coronal optical depth reported here and in \citealt{N11a} and
interpreted as a result of mass ejection processes). Presently, it is
unclear if this difference is due to their wider energy coverage
($\sim4\times$), our superior phase resolution ($\sim10\times$), or
differences in our models, i.e.\ {\tt simpl} versus {\tt compPS}
(which includes non-thermal electrons). But the similarities in our
results are highly encouraging, and there is hope that new theoretical 
studies will allow us to converge on a complete understanding of the
$\rho$ state in the near future.

\section{CONCLUSIONS} 
GRS 1915+105 exhibits a fascinating array of variability in its X-ray
lightcurve. It is believed that many of its variability classes are
related to limit cycles of accretion, advection, and ejection, but 
the origin of these variability classes is only understood to the
degree that we understand their diversity. This is particularly true
for the $\rho$ state, which is similar to several other variability
classes, and which displays striking variations from observation to
observation, particularly with respect to the number of peaks per
cycle in the X-ray lightcurve (e.g. \citealt{Massaro10}; this work). 

In a previous paper \citep{N11a}, we performed a detailed,
phase-resolved spectral and timing study of GRS 1915+105 in a
double-peaked instance of the $\rho$ state. Building on our knowledge
of this type of oscillation, we have presented here a comprehensive
study of archival \textit{RXTE} observations of the $\rho$-like
cycles. Broadly, our results indicate that the variation in the 
heartbeat state lightcurves can be explained in the context of two
main pulses, one soft and one hard, whose relative strengths,
duration, and separation can vary smoothly over time. This explanation
appears to apply whether each cycle exhibits one pulse, two pulses, or
some intermediate morphology. Furthermore, phase-resolved spectra from
the \textit{RXTE} PCA clearly indicate that oscillations with very
different profiles exhibit remarkably similar spectral evolution. We
conclude that a common set of dominant accretion processes produces
the diverse lightcurves.

In this context, what is particularly remarkable about the heartbeat
state is the complex interaction between the disk, its mass accretion
rate, and the radiation it produces. The oscillation is a process set
in motion by a high external accretion rate, in which an unstable disk
periodically sends waves of mass inwards
\citep[e.g.][]{TCS97,Nayakshin00,JC05}. Responding to the increasing 
influx of matter, the disk produces enough radiation to slowly expel
some fraction of the inflow, driving the inner edge of the disk away
from the black hole. But eventually, the disk's ability to compensate
is overwhelmed; when the accretion rate remains high, the excess
matter in the inner disk plunges inwards as the temperature and
viscosity spike. But even in these final moments of the cycle,
radiation pressure still manages to eject some mass from the
disk. These are the common, interconnected processes that lead to the
$\rho$-like cycles in GRS 1915+105. 

But there is another process, the accretion disk wind, whose ubiquity
has yet to be confirmed (see our upcoming \textit{Chandra, RXTE,
  Gemini,} and \textit{EVLA} study; Neilsen et al., in 
preparation). In \citet{N11a} we showed that the bright X-ray pulses
from the inflow drive a massive, ionized wind off the outer disk
($R\lesssim10^{11}$ cm). This wind can drain as much as 95\% of the 
external mass supply from the accretion flow, suggesting that the
companion star is actually supplying matter at a rate well in excess
of Eddington (since matter reaches the black hole at roughly the
Eddington rate). Massive winds have also been seen in other black
holes (e.g.\ \citealt{King11,Ponti12} and references therein) and
neutron stars (e.g.\ GX 13+1; \citealt{U04}). Eventually, this massive
wind depletes the outer disk, creating a mass deficit. The heartbeat
oscillations continue unabated until, days or weeks later, this 
deficit propagates inwards (\citealt{Shields86,Luketic10,N11a}),
simultaneously suppressing the thermal-viscous instability, the local
Eddington effect, radiation-pressure-driven mass ejection, and the
accretion disk wind itself.  

We end on an exciting note, for it is finally clear that GRS 1915+105
is not alone. In a new outburst detected by \textit{Swift/BAT}
\citep{Krimm11} and followed up with frequent \textit{RXTE} PCA
observations, the black hole candidate IGR J17091--3624 was discovered
to exhibit heartbeats and several other of GRS 1915+105's variability
classes in its X-ray lightcurve (e.g.\ \citealt*{Pahari11};
\citealt{Altamirano11}). But if GRS 1915+105's behavior is made
possible by a high mass supply rate from its subgiant companion
\citep*{G01}, is the presence of two active ``GRS 1915+105''s in our
Galaxy consistent with models of stellar and binary evolution? Thus,
although at a much lower X-ray flux, future studies of IGR 17091--3624
will place strong constraints on the origin of accretion instabilities
like those in GRS 1915+105.\vspace{-3mm}

\acknowledgements We thank Diego Altamirano for provocative
discussions of IGR 17091--3624 and GRS 1915+105. J.N.\ gratefully
acknowledges funding support from Chandra grant G07-8044X and the
Harvard University Graduate School of Arts and Sciences, as well as
support from the National Aeronautics and Space Administration through
the Smithsonian Astrophysical Observatory contract SV3-73016 to MIT
for support of the \textit{Chandra} X-ray Center, which is operated by
the Smithsonian Astrophysical Observatory for and on behalf of the
National Aeronautics Space Administration under contract NAS8-03060.  
R.A.R.\ acknowledges partial support from the NASA contract to MIT for
the support of \textit{RXTE} instruments. This research has made use
of data obtained from the High Energy Astrophysics Science Archive
Research Center (HEASARC), provided by NASA's Goddard Space Flight
Center.

\bibliographystyle{hapj}
\bibliography{ms}

\label{lastpage}

\end{document}